\documentclass[%
 reprint,
 superscriptaddress, 
 amsmath,amssymb,
 aps,
 pra,
]{revtex4-2}

\usepackage{graphicx}
\usepackage{dcolumn}
\usepackage{bm}
\usepackage{braket}
\usepackage{comment}
\usepackage{hyperref}
\usepackage{xcolor}

\begin{document}

\title{Simulating generic single-qubit open-dynamics via polarization-frequency coupling in a photonic interferometer}

\author{Kalle Raikisto}
\email{kalle.v.raikisto@utu.fi}
\affiliation{Department of Physics and Astronomy, University of Turku, FI-20014 Turun yliopisto, Finland}

\author{Alberto Ferrara}
\affiliation{Dipartimento di Ingegneria, Universit\`{a} degli Studi di Palermo, Viale delle Scienze, 90128 Palermo, Italy}

\author{Tom Kuusela}
\affiliation{Department of Physics and Astronomy, University of Turku, FI-20014 Turun yliopisto, Finland}

\author{Rosario Lo Franco}
\affiliation{Dipartimento di Ingegneria, Universit\`{a} degli Studi di Palermo, Viale delle Scienze, 90128 Palermo, Italy}

\author{Jyrki Piilo}
\affiliation{Department of Physics and Astronomy, University of Turku, FI-20014 Turun yliopisto, Finland}

\date{\today}

\begin{abstract}
We propose a photonic platform for simulating arbitrary single-qubit open-system dynamics using a single photon in an open Mach–Zehnder interferometer. A birefringent quartz plate induces a coupling between the polarization and frequency degrees of freedom. By treating the latter as an effective environment, we analytically derive the reduced polarization dynamics. We show that the resulting evolution is characterized by a controllable interplay between populations and coherence, instead of the usual dephasing caused by quartz plates. By adjusting the photon frequency distribution and interferometric parameters, we demonstrate that target single-qubit states can be efficiently reproduced through a tunable optical protocol expected to work under accessible experimental conditions. The simulator is benchmarked against paradigmatic open-system evolutions, including depolarization and non-Markovian dynamics, achieving high accuracy. Our results establish polarization-frequency engineered photonic interferometers as a versatile protocol for simulation of open quantum systems.
\end{abstract}

\maketitle

\section{Introduction}

Open quantum systems interact with their environment, thus undergoing a decoherence process. This dynamical process can be classified to be either Markovian or non-Markovian depending on whether the system experiences memory effects \cite{rivas2014quantum,breuer2016colloquium}. Markovian dynamics are well studied due to their simpler mathematical structure, but non-Markovian dynamics have seen an increase in interest due to a better understanding of the nature of memory effects \cite{breuer2009measure, rivas2010entanglement, pollock2018non} and improvements in mathematical and numerical methods \cite{chruscinski2016non, pollock2018non, cerrillo2014non, settimo2026quantum, suess2014hierarchy}. This has led to a new growing research direction treating them as a resource \cite{bylicka2014non, berk2021resource}. Non-Markovianity has been used in numerous applications in quantum technology such as quantum error mitigation \cite{taranto2021non, hakoshima2021relationship, wang2025non, endo2025non}, decoherence control \cite{white2020demonstration, white2022characterization}, quantum communication \cite{bylicka2014non, li2019non}, entanglement preservation \cite{bellomo2007non, shi2016entanglement}, quantum metrology \cite{chin2012quantum, altherr2021quantum, yang2019memory, yang2024control}, and quantum cryptography \cite{vasile2011continuous, utagi2020ping, sabogal2025decoherence}.

The ability to control and engineer non-Markovian dynamics experimentally is thereby important to realize these applications. 
Subsequently, the development of simulators for open quantum systems have gained increasing interest both for producing the desired non-Markovian dynamics for applications and also as a means to further understand the nature of memory effects. 

Indeed, the last 15 years have seen an increase in the number of different simulator setups. 
While there have been various experimental platforms, such as nuclear magnetic resonance \cite{bernardes2016high, khurana2019experimental}, trapped ions \cite{wittemer2018measurement}, multi-qubit conducting devices \cite{white2020demonstration}, and nitrogen-vacancy centers in diamonds \cite{haase2018controllable, dong2018non}, photonic simulators have remained useful and common choice for this task.

One of the first optical simulators was a Sagnac interferometer combined with half- and quarter-wave plates at different angles to produce several different decoherence channels \cite{salles2008experimental}, and similar setups have been studied since \cite{chiuri2012linear, cuevas2019all}.  
While other proposals -- sometimes based on very different approaches --  have been made \cite{cialdi2017all, rossi2017non, uriri2020experimental, orieux2015experimental, zou2013photonic, jin2015all}, there is perhaps no consensus on the best approach, yet, and the question of generality still requires attention.

Among these, a common approach for simulating decoherence in optical qubits is by introducing a birefringent medium. The birefringent medium creates an interaction between the system (polarization) and an effective environment 
(frequency), and this polarization-frequency coupling results in dephasing. This method was first introduced in \cite{liu2011experimental} and has since been used extensively to study open quantum system 
dynamics \cite{tang2012measuring, liu2018experimental, liu2024, lyyra2020experimental, laine2014nonlocal, liu2020experimental, liu2016efficient, hamedani2020photonic, siltanen2020distributing, laine2013erratum, liu2013photonic}.

While the birefringent medium has been used to create a qubit dephasing simulator \cite{liu2018experimental}, it has been so far limited to this specific kind of noise. 
In this article, we expand this protocol  
from dephasing to general decoherence by placing the birefringent medium inside a Mach-Zehnder interferometer and exploiting also interference in addition to dephasing.
This combination has been previously shown to produce non-Markovian dynamics 
\cite{siltanen2021interferometric}, but here we demonstrate that it can be used to create a general qubit decoherence simulator.
Our simulator allows the simulation of almost any qubit open quantum system dynamics, which we demonstrate using two paradigmatic examples: Depolarizing qubit channel and non-Markovian dissipating qubit. 

The paper is organized as follows. In Sec.~\ref{chap:1} we present the optical setup and explicitly derive the single photon state after the interferometric process. Sec.~\ref{chap:2} is devoted to a more detailed description of the output state properties, assuming a specific form for the frequency distribution. Afterwards, in Sec.~\ref{chap:3}, we present a protocol to simulate a generic single-qubit evolution through the interferometric setup. Finally, in Sec.~\ref{chap: concl} we draw conclusions and discuss some outlooks.

\begin{figure}
    \includegraphics[width=0.50\textwidth]{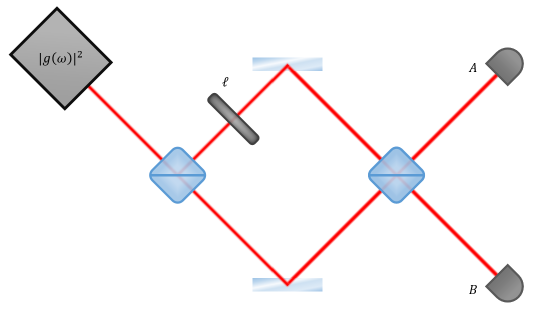}
    \caption{\label{setup}Diagram of the setup. The system is a Mach-Zehnder interferometer consisting of 2 beam splitters (blue), where the photon goes through quartz plate (grey). The initial state is determined by the choice of frequency distribution $|g(\omega)|^2$ and the thickness of the quartz plate, as well as by its initial probability amplitudes $C_H$ and $C_V$ and relative phase $\phi$. }
\end{figure}

\section{Dynamics of polarization state}
\label{chap:1}
The system we study, as shown in Fig.~\ref{setup}, consists of a single photon going through an open Mach-Zehnder interferometer. We characterize the photon state via a binary degree of freedom represented by the photon polarization $(\ket{H}, \ket{V})$ and via its frequency $(\ket{\omega}, \,\, \omega \in \mathbb{R})$, whose continuous probability amplitude distribution is described by a function $g(\omega)$. 
Under this framework, the initial state can be expressed as:
\begin{align} \label{eq:init_state}
    \ket{\psi(0)} &= C_H \ket{H} \int d\omega \,g(\omega) \ket{\omega} \notag\\
    &+ e^{i \phi} C_V \ket{V} \int d\omega \, g(\omega) \ket{\omega} . 
\end{align}

Here, $C_H$ and $C_V$ represent the initial populations in the two polarization states. Without loss of generality, we pick $C_{H}, C_V \in \mathbb{R}$, with a generic phase difference $\phi$. After being initialized in this state, the photon enters the interferometer. The full state entering into the beam splitter can be formally expressed as $\ket{\psi(0)}_A\otimes\ket{0}_B$. Here, the label $A$ and $B$ refer to the two distinct paths of the interferometer, and $\ket{0}$ represent a vacuum state.
Then, on one of the two paths, a quartz plate induces dephasing on the polarization qubit. As the thickness of such plate $\ell$ can be experimentally set, this process can be controlled. Finally, the photon state over the two branches enters another beamsplitter, where interference occurs between the two paths. The final state of the photon is then detected at the output ports of the interferometer, where we can measure the polarization state of the photon.
To track the effect of the interferometer, we determine the state of the system step by step. Beginning from the initial state in
Eq.~(\ref{eq:init_state}) we first calculate the state of the photon after it goes through the first beam splitter and the quartz plate. The evolution induced by the latter is described through the action of a unitary operator acting on the frequency degrees of freedom as $U_{\lambda} \ket{\omega}= e^{i \omega n_\lambda \ell/c} \ket{\omega}$, where the speed of light will be hereafter set to $c=1$. This means that length $\ell$ will be measured in seconds spent inside the quartz plate, while the frequency $\omega$ will remain in the units of hertz. Here, $n_\lambda$ is the refractive index, which depends on the polarization $\lambda$. If we operate with this on the initial state, we get the total state of the photon
\begin{equation}
        \ket{\Psi} = \frac{1}{\sqrt{2}}\big[U_{\lambda}\ket{\psi (0)}_A \ket{0}_B + \ket{0}_A \ket{\psi(0)}_B\big].
\end{equation}
The density matrix of the polarization state after the photon has gone through the quartz plate can be calculated by taking a partial trace over the frequency degree of freedom of the photon state in path $A$. The dephasing caused by the quartz plate will result in the following state:

\begin{align}\label{rho}
    \rho_{in} = \begin{pmatrix}
|C_H|^2 & C_H C_V \kappa_{in} \\
C_H C_V \kappa_{in}^* & |C_V|^2\\
\end{pmatrix}, 
\end{align}
From this equation, we see that the evolution caused by the quartz plate only affects the off-diagonal elements, leaving the populations unchanged. The decoherence function here is
\begin{align}
\kappa_{in} = \int d\omega |g(\omega)|^2e^{i \omega \Delta n l} e^{i \phi}, 
\end{align}
where the difference in the indices of refraction of the two polarization components is given by $\Delta n = n_H - n_V$. For a fixed quartz plate thickness, the decoherence function can be controlled by appropriately changing the initial state frequency distribution $|g(\omega)|^2$. An extremely narrow frequency distribution will result in a simple phase shift. Conversely, a broad $|g(\omega)|^2$ will reduce the absolute value of the coherence, causing dephasing. After this step, the second beam splitter results in interference between the states on the two paths, with one containing the aforementioned quartz plate. This produces the following dynamics for the qubit polarization state:

\begin{align}\label{eq: rho_out_j}
    \rho_j = \frac{1}{4p_j}\begin{pmatrix}
(2 + (-1)^j \eta_H)|C_H|^2 & C_H C_V \kappa_j \\
C_H C_V \kappa_j^* & (2 + (-1)^j \eta_V)|C_V|^2\\
\end{pmatrix}, 
\end{align}
where $j \in \{A, B\}$ labels the two paths of the interferometer. In the exponents of the diagonal elements, path $A$ ($B$) corresponds to $0$ ($1$). For more details of the derivation, see Appendix \ref{details}. In Eq.~(\ref{eq: rho_out_j}), we have introduced the notation

\begin{align}\label{eq:eta}
    \eta_\lambda := \int d\omega 2|g(\omega)|^2 cos(\omega n_\lambda \ell).
\end{align}
The probabilities for the paths are 

\begin{align}\label{eq:path_prob}
    p_j = \frac{2+(-1)^j|C_H|^2\eta_H + (-1)^j |C_V|^2 \eta_V}{4},
\end{align}
and on the off-diagonal, we have a decoherence function $\kappa_j$ that depends on the path as well, given by

\begin{align}\label{eq:decoherence_outside}
     \kappa_j&= e^{i\phi} \int d\omega |g(\omega)|^2 \notag \\
    &\cdot (1 + e^{i \omega \Delta n \ell}   +(-1)^j e^{i \omega n_H \ell} +(-1)^j e^{-i \omega n_V \ell}).
\end{align}

On the diagonal of the state shown in Eq.~(\ref{eq: rho_out_j}), we see rapid oscillations in $\eta_\lambda$ caused by interference. This result reveals an interesting dynamical structure: we do not have a simple dephasing process where only the off-diagonal terms evolve, nor do we have evolution that is completely determined by the populations on the diagonal, rather it is a combination of both. In this scenario, the initial frequency distribution $|g(\omega)|^2$ affects both the diagonal and the off-diagonal, acting as a spectral function of the effective environment.
Concurrently, $\ell$ influences both the populations and the coherence, due to the fact that the quartz plate induces dephasing on the polarization state before the action of the second beam splitter. This enables a non-trivial interference process, which will be described in more detail in the following section. Additionally, we investigate how the elements of the output reduced density matrix can be controlled by tuning the frequency distribution.
\begin{figure*}
    \centering
    \includegraphics[width=0.99\linewidth]{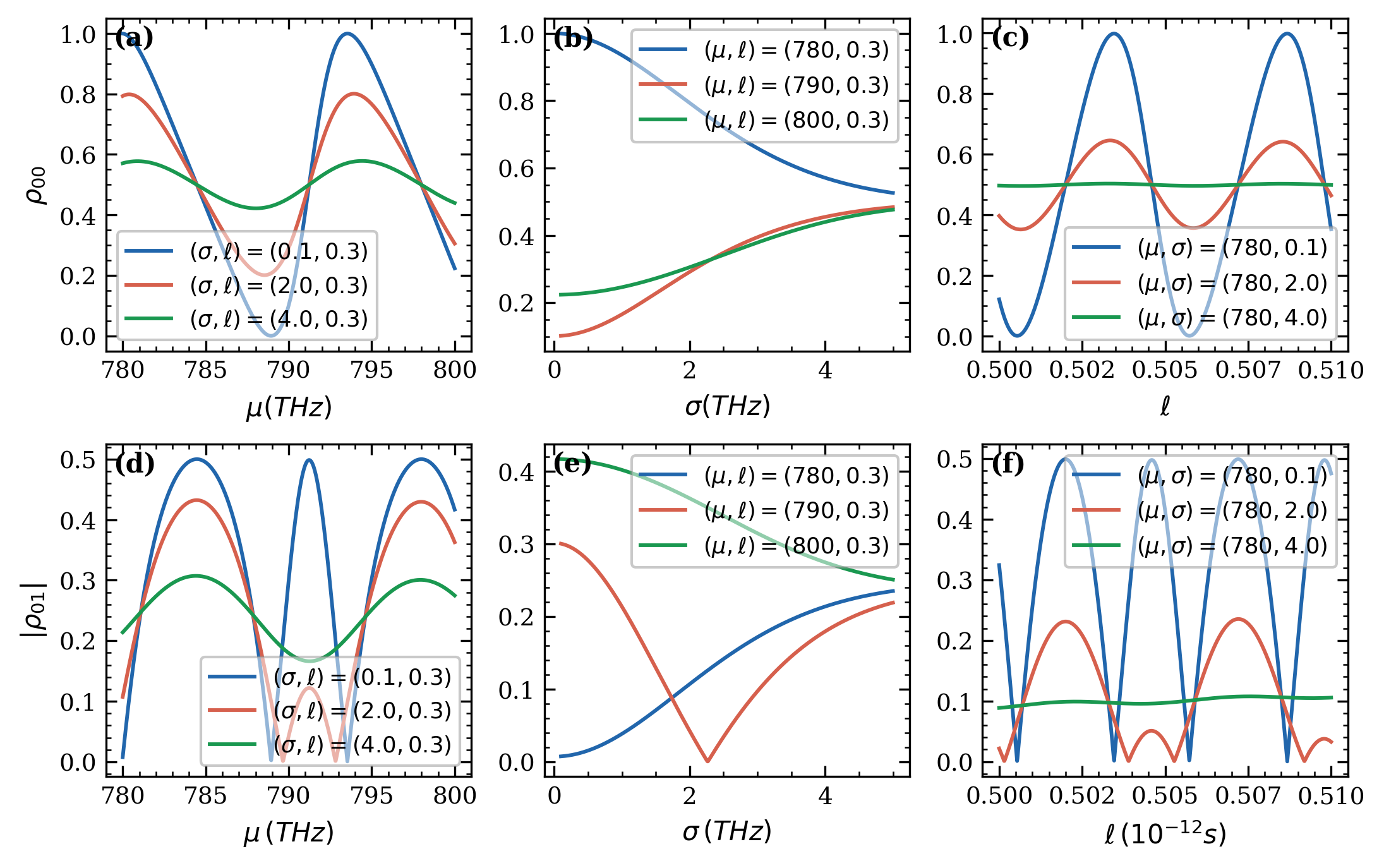}
    \caption{The diagonal and off-diagonal values of the polarization state for one of the output branch of the interferometer, as a function of the controllable parameters $\mu, \sigma, \ell$. The other parameters are set to $C_H = 0.5$, $n_H = 1.553$ and $n_V = 1.544$. \textbf{Panels (a, b, c):} $\rho_{00}$ as function of $\mu$, $\sigma$ and $\ell$ respectively. \textbf{Panels (d, e, f):} $|\rho_{01}|$ as function of $\mu$, $\sigma$ and $\ell$ respectively.}
    \label{MZ_parameters}
\end{figure*}

\section{Characterization of the Controllable Output State}
\label{chap:2}
For the purpose of creating a general qubit decoherence simulator, we consider $\rho_A$, the polarization state detected at the end of path $A$. In the following, we set the frequency distribution to be a Gaussian, such that $|g(\omega)|^2 = \frac{1}{\sigma\sqrt{2 \pi}}e^{-\frac{(\omega - \mu)^2}{2 \sigma^2}}$, with the constraint that $\mu \gg \sigma$, so that the mathematical contribution due to negative frequencies is negligible. Since the dynamics is controlled by the distribution function and the length of the quartz plate, this choice leaves us with a set of six experimentally controllable parameters $\{\mu, \sigma, \ell, C_H, \phi, \Delta n\}$, enabling a variety of dynamical behaviors that allow us to simulate a generic qubit dynamics. Under this assumption, the integrals in Eqs.~\ref{eq:eta} and \ref{eq:decoherence_outside} can be explicitly evaluated, resulting in:

\begin{align}
    &\eta_\lambda(\mu, \sigma, \ell) = 2 \cos{(\mu n_\lambda \ell) e^{- \frac{(n_H \ell \sigma)^2}{2}}} \label{eq:eta_exact} \\
    &\kappa_A(\mu, \sigma, \ell, \phi) = e^{i\phi}\cdot \bigg[1 + e^{i \mu \Delta n  \ell - \frac{1}{2} (\Delta n \sigma \ell)^2} \notag \\
        &+ e^{i \mu n_H \ell - \frac{1}{2} (n_H \sigma \ell)^2} + e^{i \mu n_V \ell - \frac{1}{2} (-n_V \sigma \ell)^2} \bigg].\label{eq:kappa_exact}
\end{align}
We assume to keep $n_H = 1.553$ and $n_V = 1.544$, matching the natural properties of quartz and fixing the value of $\Delta n$.
Let us now take a closer look at the remaining five parameters and how they affect the polarization qubit, namely $\{\mu, \sigma, \ell, C_H, \phi \}$. The initial phase difference $\phi$ of Eq.~(\ref{eq:init_state}) directly changes the phase of the off-diagonal elements and gives us full control over their values. Practically, this can be done by setting an initial constant phase between the horizontal and the vertical polarization. The initial probability amplitude $C_H$ linearly and quadratically rescales the coherence and the population, respectively. The behavior induced by the other three parameters is shown in Fig. \ref{MZ_parameters}. Panel (a) displays the periodic oscillations of the diagonal elements as a function of $\mu$. Similar dynamics can be seen in panel (d) where the off-diagonal element is depicted as a function of $\mu$. These oscillations span almost the entire range of the diagonal values, although the amplitude of the oscillations depends on the values of the other parameters $\ell$ and $\sigma$. The latter, in turn, does not produce oscillations on the diagonal; instead, it controls their amplitude. In panels (b, e), we see that $\rho_{00}$ and $|\rho_{10}|$ tend to a constant value for increasing $\sigma$. Interestingly, a high oscillation amplitude for both the diagonal and off diagonal values at varying $\mu$ is achievable when $\sigma$ is small (i.e. $ \sigma \approx0.1$ THz). This can be understood in light of the interplay mechanism between dephasing and interference, which induces a tunable polarization-dependent routing between the two output ports of the interferometer, as described by the term $[2 + (-1)^j \eta_\lambda]/p_j$. The constant contribution $\sim 2/p_j$ can be interpreted as a uniform baseline distribution; in total absence of interference, as shown in Eq.~(\ref{eq: rho_out_j}), this would result in $\frac{1}{2}$, meaning that (i) both output ports would share the probabilities equally ($p_j=\frac{1}{2}$) and (ii) the population ratio $C_H/C_V$ would remain untouched. Subsequently, $(-1)^j$ accounts for the conservation of probability among the two distinct outputs of the beamsplitter. This implies that whenever one port $A$ or $B$ experiences constructive interference, the other must undergo destructive interference. Finally, the function $\eta_\lambda$ can be interpreted as an interference visibility at the level of the second beamsplitter. If the frequency distribution is extremely narrow (i.e. $\sigma \to 0, |g(\omega)|^2 \to \delta(\omega - \mu)$), dephasing due to the quartz plate is inhibited $(|\kappa_{in}|\to1)$ and interference is maximized, with $\eta_\lambda \to 2 \cos(\mu n_\lambda\ell)$. Conversely, for a wide frequency spread $\sigma$, the same decoherence function goes to its minimum value $\kappa_{in} \to 0$. In this case, the photon state on the path affected by the quartz plate becomes a classical statistical mixture of horizontal and vertical polarization. Such a state does not result in an interference phenomenon on the diagonal elements of the $\rho_j$.

A complementary interpretation can be given within the framework of open-quantum systems. Since the frequency degrees of freedom behave as an effective environment, the quartz plate entangle the polarization with distinct frequencies as $\ket{\lambda}\ket{\omega}$ evolves into $e^{i\omega n_\lambda \ell} \ket{\lambda}\ket{\omega}$. In this process, some \textit{which-polarization} information is effectively leaked into the environment and lost when taking the trace, reducing the degree of indistinguishability between the photon state on path $A$ with respect to the photon state in path $B$. This, in turn, decreases the interference visibility.

\section{Single qubit dynamics simulator}
\label{chap:3}
Having investigated the properties of the output state in our interferometric setup, we can now show how a generic dynamical evolution of a qubit can be simulated via a controlled tuning of dephasing and interference. The state of a generic target qubit can be described through its associated density matrix as
\begin{equation}
    \rho (t) = \begin{pmatrix} \rho_{00}(t) & e^{i \beta(t)}|\rho_{01}|(t) \\ e^{-i \beta(t)}|\rho_{01}|(t) & 1-\rho_{00}(t) \end{pmatrix}.
\end{equation}
This implies that, by exploiting the physical properties of quantum states, we can define and characterize a qubit state, be it pure or mixed, through three distinct real numbers $ \rho_{00}$, $|\rho_{01}|^2$, and $\beta$. Additionally, due to the positivity requirement of density matrices, we have $|\rho_{00}(1-\rho_{00})|>|\rho_{01}|$. To simulate this qubit state, we have to match it to our output polarization state, as given in Eq.~(\ref{eq: rho_out_j}). Once again, we make the same assumptions about our interferometric setup as in Sec.~\ref{chap:2} and use the exact solutions shown in Eqs.~\ref{eq:eta_exact},~\ref{eq:kappa_exact}. Then, by direct comparison, we are left with a system of three algebraic equations: 

\begin{equation}\label{eq: optim_cond_general}
    \left\{ 
    \begin{array}{l}
        |C_H|^2 \left[\frac{2 + \eta_H}{4 p_A}\right] = \rho_{00}(t)\\
        \frac{C_H C_V}{\alpha}\frac{|\kappa_A|}{4p_A} = |\rho_{01}| \\
         \phi = + \beta(t) - \delta. \\
    \end{array} 
\right.
\end{equation}
Here, we have expressed the decoherence function in the form $\kappa_A = e^{i \delta} |\kappa_A|$ to explicitly separate the conditions for the absolute value and the phase of the coherence. As previously stated, $\phi$ only affects the phase of the off-diagonal elements. Therefore, the last equation is trivially satisfied by appropriately tuning the constant phase $\phi$. That leaves us with four parameters that we can experimentally control in our interferometric setup: Initial populations $C_H$, length of the quartz plate $\ell$, and the mean and variance of the frequency distribution $\mu$, $\sigma$. Using these, we can manipulate the polarization state of the photon to closely match the \textit{target} qubit undergoing general dynamics of interest at a specific time $t$. Optimizing all such parameters for each point of the simulation would, of course, yield the most accurate theoretical simulation results. However, to keep the simulator practical for experimental use, we pick only a single parameter to be optimized. From figure Fig.~\ref{MZ_parameters} we can see that all of the parameters affect both diagonal and off-diagonal values, but some are clearly better solutions than others. We choose $\mu$ since it provides a wide range of possible diagonal and off-diagonal values and can be more easily controlled experimentally using filters, whereas the length of the quartz plate can be more inconvenient to control exactly. This is relevant, as the present protocol requires the tuning parameter to be set for each desired $t$ in the simulation. The rest of the parameters $\{\sigma, \ell, C_H\}$ are set to be constant throughout the whole simulation, although optimized values for them are initially scouted to appropriately enlarge the achievable value range. These parameters are found using a combination of grid search and differential evolution. Then, for each time $t$ we want to simulate, a new value for $\mu$ is selected. For its tuning process, we use a joint optimization algorithm consisting of a rough grid search that finds us a set of initial guesses. Finally, starting from these initial guesses, we optimize using the Limited-memory Broyden–Fletcher–Goldfarb–Shanno algorithm \cite{byrd1995limited} and pick the best result. 

To showcase the effectiveness of the protocol, we put it to the test with two specific but paradigmatic instances of single qubit dynamics: (i) a depolarizing channel and (ii) non-Markovian dynamics manifesting coherence revivals. 

\subsection{Depolarizing channel}

\begin{figure}
    \centering
    \includegraphics[width=0.99\linewidth]{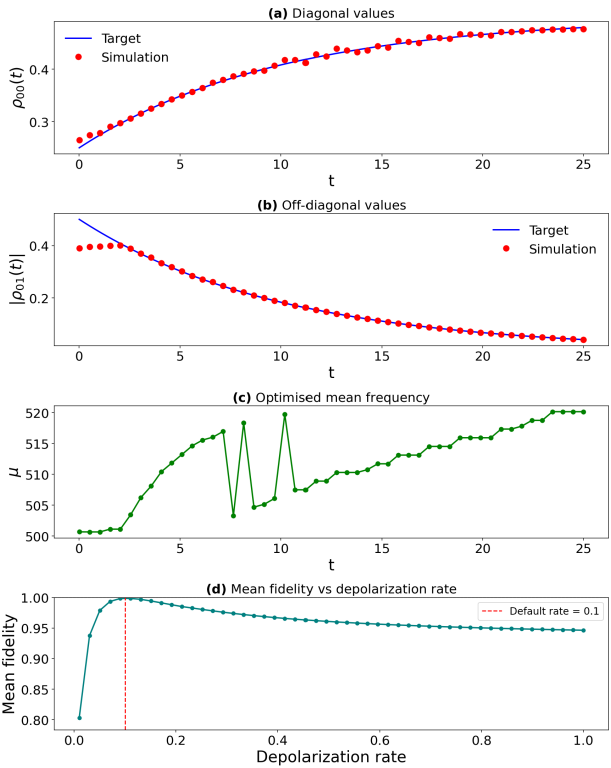}
    \caption{Results of the single qubit simulation for the depolarizing channel, as described in Eq.~(\ref{eq:depolarization}) \textbf{(a):} Comparison of the diagonal element $\rho_{00}$ \textbf{(b):} Comparison of the off-diagonal coherence \textbf{(c):} Values of the mean frequency $\mu$ in the unit of $THz$. \textbf{(d):} Mean fidelity of the simulation shown as function of the depolarization rate $\gamma$. For \textbf{(a)-(c)} the parameters are $\ell=8.688229 \cdot 10^{-12}~s$, $\sigma=0.1 ~THz$, $n_H = 1.553$, $n_V = 1.544$, $|C_H| = 0.527694$, $\gamma = 0.1$. For \textbf{(d)} parameters $\ell$, $C_H$ and $\sigma$ are kept while a new set of $\mu$ values is separately optimized each $\gamma$.}
    \label{depolarization}
\end{figure}

Depolarizing noise is a quantum channel that models the process by which a qubit state is progressively degraded into the maximally mixed state \cite{nielsen2010quantum}. Physically, it represents the worst-case scenario of isotropic noise, where information is lost uniformly in all directions. Under this decoherence process, the Bloch vector associated with a given quantum state progressively shrinks, losing coherence and information on the populations. By introducing a timescale $\gamma$ for the decoherence process, the evolution of a generic qubit under depolarizing noise can be expressed as
\begin{equation} \label{eq:depolarization}
    \rho(t) = \rho (0)e^{-\gamma t} + (1 - e^{-\gamma t}) \frac{\mathbf{I}}{2},
\end{equation}
where $\mathbf{I}$ is the $2$x$2$ identity matrix. 

In Fig.~\ref{depolarization} the simulation is compared to exact dynamics of Eq.~(\ref{eq:depolarization}) and the optimized parameter values for $\mu$ returned by the protocol are plotted. The simulation is shown to be accurate as long as the values are within the range of the simulator. The target qubit's initial state is a quantum superposition of $(\ket{0}$ and $\ket{1})$, where we encode the qubit state $\ket{0}$ as $\ket{H}$ and $\ket{1}$ as $\ket{V}$. We tested with a range of different initial states and depolarization rates and our protocol was able to successfully simulate them. This is shown in Fig.~\ref{depolarization} (d) where the fidelity between the simulated state and the target is calculated as a function of the depolarization rate $\gamma$. Parameters $C_H, \ell, \sigma$ are kept the same for each $\gamma$ and the mean fidelity shown is calculated over 50 points in time $t$. We can see that while the best fidelity is found for the default rate for which the interferometer parameters were optimized for, there is significant flexibility. Even when altering the simulated dynamical process, the fidelity remains quite good, although the parameter values are not specifically chosen for that process. This means that the constant parameters do not need to be perfectly optimized for the fidelity of the simulator to be excellent. Imperfections in these parameters can be made up for with adjustments in the choice of the $\mu$ values.

\subsection{Non-Markovian dynamics}
Another paradigmatic example of non-trivial dynamical behavior is offered by dissipative non-Markovian evolution in the context of open quantum systems. 
As a showcase of this phenomenon, we consider a qubit that can exchange excitations with a structured bosonic bath, characterized by multiple modes. Under the rotating wave approximation, the full Hamiltonian of the system can be written as \cite{breuer1999stochastic}
\begin{equation}
    \begin{split}
            \hat{H} & = \frac{\omega}{2} \hat{\sigma}_z + \sum_k \omega_k \hat{a}^\dagger_k\hat{a}_k  +\sum_k (g_k\hat{\sigma}_+ \hat{a}_k + h.c.)
    \end{split}
\end{equation}
Here, $\hat{\sigma}_z = \{(1,0),(0,-1)\}$ represents a pauli matrix while $\hat{\sigma}_+$ is the qubit ladder operator. $\hat{a}_k$ and $\hat{a}^\dagger_k$ are, respectively, the annihilation and creation operators for environment mode $k$. 
Physically, this system can be mapped to a two-level emitter in a lossy cavity. According to the features of the system-environment coupling distribution represented by the set of parameters $g_k$, non-Markovian dynamics can be observed in the form of repeated coherence revivals when the initial state is in a quantum superposition, due to an information backflow between the system and environment.
The time evolution of a qubit in this scenario can be analytically derived \cite{breuer1999stochastic, breuer2016colloquium}, resulting in
\begin{align}
    & \rho_{11}(t) = |G(t)|^2 \rho_{11}(t_0=0) \\
    & \rho_{01}(t) = G(t) \rho_{01}(t_0=0).
    \label{eq: NM dynamics}
\end{align}
The reduced dynamics of the qubit can also be cast in a master equation form
\begin{equation}
    \begin{split}
            \dot{\hat{\rho}}(t) = & - \frac{i}{4}S(t)[\hat{\sigma}_z, \hat{\rho}(t)]\\
        &+ \gamma(t)\left[ \hat{\sigma}_- \hat{\rho}(t) \hat{\sigma}_+ - \frac{1}{2}\{ \hat{\sigma}_+\hat{\sigma}_-, \hat{\rho}(t)\}\right].
    \end{split}
\end{equation}
The full expression for the time-dependent Lamb shift $S(t)$ as well as the decay rate $\gamma(t)$ and the decoherence function $G(t)$ are reported in Appendix~\ref{app: NME}.
\begin{figure}
    \centering
    \includegraphics[width=0.99\linewidth]{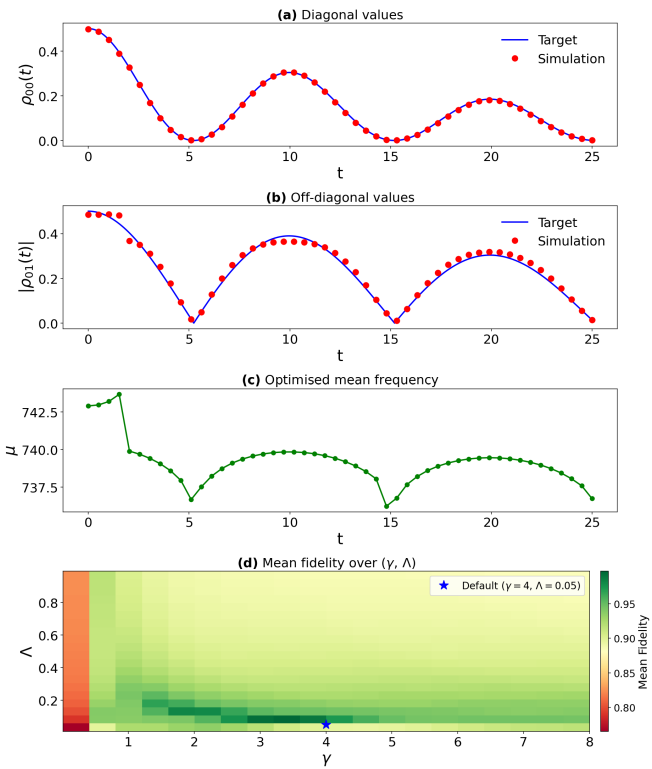}
    \caption{Results of the single qubit simulation for a non-Markovian dynamics, as described in Eq.~(\ref{eq: NM dynamics}) \textbf{(a):} Comparison of the diagonal element $\rho_{00}$ \textbf{(b):} Comparison of the off-diagonal coherence \textbf{(c):} Values of the mean frequency $\mu$ in the unit of $THz$. \textbf{(d):} Colormap of the mean fidelity as a function of $\gamma$ and $\lambda$. For \textbf{(a)-(c)} the parameters are $\ell =0.101579 \cdot 10^{-12} ~s$, $\sigma=0.637 ~THz$, $n_H = 1.553$, $n_V = 1.544$, $|C_H| = 0.3$, $\gamma = 4$, $\lambda = 0.05$. For \textbf{(d)} parameters $\ell$, $C_H$ and $\sigma$ are kept while a new set of $\mu$ values is separately optimized each $\gamma, \lambda$. }
    \label{fig: NM_sim}
\end{figure}

In Fig.~\ref{fig: NM_sim}, we show a comparison between the evolution of the qubit in the non-Markovian regime, together with the polarization degrees of freedom of the photon coming out of the interferometer for different choices of $\mu$ at each time $t$. The target qubit's initial state is once again $\ket{\phi(0)}=\frac{1}{\sqrt{2}}(\ket{0} + \ket{1})$, where we encode the qubit state $\ket{0}$ as $\ket{H}$ and $\ket{1}$ as $\ket{V}$. Different decoherence functions $G(t)$ are tested in terms of its parameters $\gamma, \lambda$ and the fidelity for each set is shown in Fig.~\ref{fig: NM_sim} (d). Interferometer parameters $C_H, \ell, \sigma$ are once again kept constant for each $\gamma, \lambda$ and the mean fidelity shown is calculated over 50 points in time $t$. The results echo the ones from the depolarization channel, with the fidelity peaking at the default point that the parameters were optimized for but remaining quite good even when the dynamical process is altered. For Markovian amplitude damping, see Appendix~\ref{M-AD}.

\section{Conclusions}
\label{chap: concl}

In this work, we have theoretically introduced and numerically benchmarked a versatile photonic platform for simulating the dynamics of single-qubit open quantum systems. The proposed scheme relies on a single photon propagating through an open Mach--Zehnder interferometer, where the polarization degree of freedom plays the role of the system qubit, while the frequency degree of freedom acts as an effective environment. A birefringent quartz plate placed in one arm of the interferometer induces a controllable polarization-frequency coupling, and the subsequent recombination at the second beam splitter converts this coupling into a nontrivial evolution of both populations and coherences. This mechanism significantly extends the standard use of birefringent media as pure-dephasing simulators, allowing one to access a much broader class of single-qubit dynamical maps.

A central result of our analysis is the explicit derivation of the reduced polarization state at the output ports of the interferometer. We have shown that the final polarization dynamics is governed by the combined action of dephasing and interference: the frequency distribution $|g(\omega)|^2$ controls the effective environmental spectral properties, while the interferometric recombination produces polarization-dependent population redistribution. As a consequence, the output state is not restricted to a phase-damping evolution, but can reproduce target qubit states characterized by independently evolving populations and coherences within the accessible range of the optical parameters. For a Gaussian frequency distribution, we have obtained closed analytical expressions for the relevant interference and decoherence functions, making transparent how the mean frequency, spectral width, quartz thickness, and initial polarization amplitudes determine the simulated state.

We have then formulated an operational protocol for simulating generic single-qubit dynamics. In order to keep the scheme experimentally practical, we selected the mean frequency $\mu$ as the main time-dependent tuning parameter, while keeping the quartz-plate thickness, spectral width, and initial polarization amplitudes fixed during a given simulation. This choice avoids the need for mechanically changing the birefringent element at every simulated time step and is compatible with standard spectral filtering techniques. Despite this constraint, the protocol retains substantial flexibility, since changes in $\mu$ can efficiently modify both the diagonal and off-diagonal elements of the output density matrix.

The performance of the simulator has been tested on paradigmatic examples of open-system dynamics. First, we considered depolarizing noise, where both populations and coherences evolve toward the maximally mixed state. The numerical results show that the interferometric protocol accurately reproduces the target dynamics over a wide range of depolarization rates, with high average fidelity even when the fixed interferometric parameters are not reoptimized for each rate. Second, we investigated a dissipative non-Markovian evolution associated with a structured bosonic environment, where coherence revivals arise from information backflow between the system and its environment. The proposed setup successfully captures these nonmonotonic features, demonstrating its ability to emulate dynamics beyond simple Markovian decay. The additional amplitude-damping example further supports the general applicability of the method to standard single-qubit noise models.

From a physical perspective, the present approach can be viewed as a form of optical reservoir engineering. The frequency degree of freedom, rather than being treated merely as an uncontrolled source of decoherence, is deliberately shaped and exploited as a programmable environment. At the same time, the interferometer transforms the information encoded in the polarization-frequency correlations into controllable changes of the reduced system state. This combination of engineered dephasing and interference provides a compact and conceptually transparent route to the simulation of open quantum dynamics using linear optical elements.

The experimental simplicity of the scheme is one of its main advantages. The required ingredients---single photons, balanced beam splitters, birefringent quartz plates, polarization preparation and analysis, and spectral filtering---are all standard tools in photonic quantum-optics laboratories. Moreover, the protocol does not require ancillary qubits or active feedback, and the relevant parameters lie within experimentally accessible regimes. These features make the setup a promising candidate for near-term demonstrations of programmable open-system dynamics.

Several directions can be envisioned for future work. A natural extension would be to move beyond single-qubit dynamics and exploit multiphoton interference to simulate multipartite open-system evolution, including correlated noise, entanglement decay, and entanglement revivals. Engineering frequency correlations between photons could also provide access to structured environments with nonlocal memory effects. In addition, combining the present interferometric architecture with tunable spectral shaping or integrated photonic components may improve scalability and enable more complex reservoir-design strategies. More broadly, the proposed platform offers a flexible testbed for studying the role of environmental memory, information backflow, and engineered decoherence in quantum technologies.

\begin{acknowledgments}
K.R. acknowledges financial support from the Quantum doctoral pilot QDoc. R.L.F. acknowledges support by MUR (Ministero dell’Università e della Ricerca) through the
PNRR Project ICON-Q - Partenariato Esteso NQSTI -
PE00000023 - Spoke 2 - CUP: J13C22000680006.
\end{acknowledgments}
\bibliography{Refs.bib}
\clearpage
\newpage

\appendix
\section{Detailed derivation of the output photon state}
\label{details}
\begin{figure}
    \centering
    \includegraphics[width=0.99\linewidth]{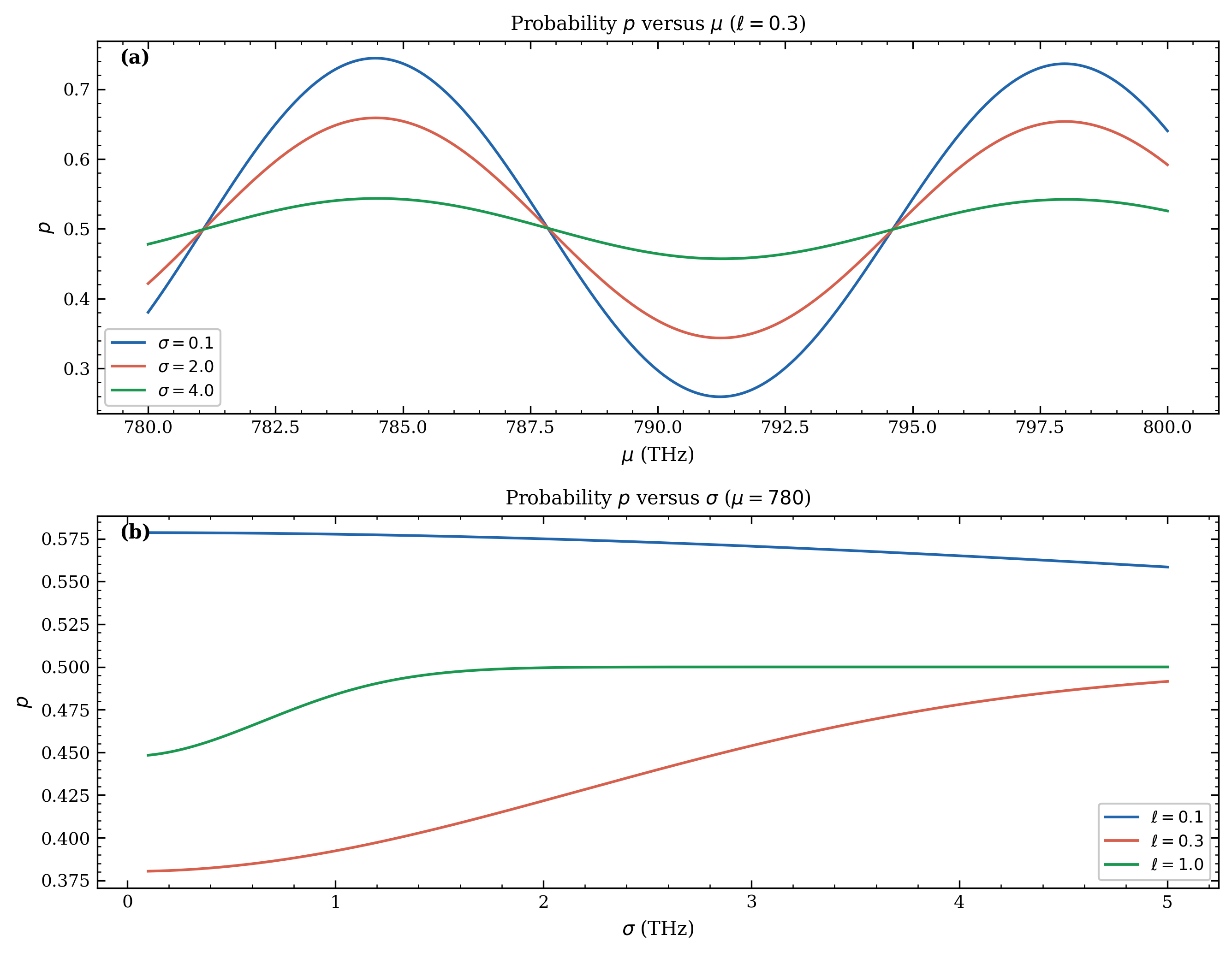}
    \caption{Probability of finding the photon in the output branch $A$ as \textbf{(a):} function of $\mu$ and \textbf{(b):} a function of $\sigma$. }
    \label{fig: prob}
\end{figure}

Here we provide a detailed derivation of the single photon output state in our interferometric setup. We start from the input state 
\begin{align}
    \ket{\psi(0)} &= C_H \ket{H} \int d\omega \,g(\omega) \ket{\omega} \notag\\
    &+ e^{i \phi} C_V \ket{V} \int d\omega \, g(\omega) \ket{\omega}. 
\end{align}
Such single photon state enters a balanced beamsplitter via path $A$. Its action can be written as
\begin{equation}
    \begin{split}
        & \ket{\phi}_A \to \frac{\ket{\phi}_A + \ket{\phi}_B}{\sqrt{2}}\\
        & \ket{\phi}_B \to \frac{\ket{\phi}_A - \ket{\phi}_B}{\sqrt{2}},
    \end{split}
\end{equation}
where $\ket{\phi}_j$ represent a generic photon state in path $j$, with $j \in \{A, B\}$ labeling one of the two paths.
After the action of the first beamsplitter, the state over both path becomes
\begin{equation}
    \begin{split}
            & \ket{\Psi_1} = \frac{1}{\sqrt{2}}\ket{\psi (0)}_A \ket{0}_B + \ket{0}_A \ket{\psi(0)}_B\big]\\
            = & \frac{1}{\sqrt{2}} \sum_\lambda C_{\lambda} e^{i \phi_\lambda}\int d\omega \; g(\omega)\big(\ket{\lambda, \omega}_A\ket{0}_B + \ket{0}_A\ket{\lambda, \omega}_B\big).
    \end{split}
\end{equation}
On branch $A$, the photon undergoes a unitary evolution due to the quartz plate, resulting in the following transformation $\ket{\lambda, \omega} \to e^{i \omega n_\lambda \ell}\ket{\lambda, \omega}$. Such a transformation can be written in terms of an interaction Hamiltonian between the polarization and the frequency $\hat{H}_I = - (n_H \ket{H}\bra{H} + n_V \ket{V}\bra{V})\int d\omega \;\omega \ket{\omega}\bra{\omega}$ \cite{siltanen2021decoherence}. We work in the assumption that the frequency intervals are small enough so that we can treat the refractive indices as constant \cite{ghosh1999dispersion}. The full state then is
\begin{equation}
    \begin{split}
            \ket{\Psi_2} & =  \frac{1}{\sqrt{2}} \sum_\lambda C_{\lambda} e^{i \phi_\lambda}\int d\omega \bigg[ g(\omega)\\
            & \times \big(e^{i\omega n_\lambda \ell}\ket{\lambda, \omega}_A\ket{0}_B + \ket{0}_A\ket{\lambda, \omega}_B\big)\bigg].
            \label{eq: psi_2}
    \end{split}
\end{equation}
At this step, by isolating the state on the first branch and taking the partial trace over the frequency degrees of freedom, we are left with
\begin{align}
    \rho_{in} = \begin{pmatrix}
|C_H|^2 & C_H C_V \kappa_{in} \\
C_H C_V \kappa_{in}^* & |C_V|^2\\
\end{pmatrix}.
\end{align}
Here, $\kappa_{in} = \int d\omega |g(\omega)|^2e^{i \omega \Delta n \ell + i\phi}$ is the first decoherence function, and corresponds to the Fourier transform of the frequency distribution. The effect of the quartz plate at this stage is limited to a coherence loss; to study the effect of the full interferometric setup, we apply the second beamsplitter to the state in Eq.~(\ref{eq: psi_2}).
\begin{equation}
    \begin{split}
            \ket{\Psi_{out}} & =  \frac{1}{\sqrt{2}} \sum_\lambda C_{\lambda} e^{i \phi_\lambda}\int d\omega \bigg[ g(\omega)\\
            & \times \big((e^{i\omega n_\lambda \ell} + 1)\ket{\lambda, \omega}_A\ket{0}_B + (e^{i\omega n_\lambda \ell} - 1)\ket{0}_A\ket{\lambda, \omega}_B\big)\bigg] \\
            & = \ket{\phi_A}_A + \ket{\phi_B}_B.
            \label{eq: psi_out} \\
    \end{split}
\end{equation}
Here, we have defined $\ket{\phi_j} = \frac{1}{2}\sum_\lambda C_\lambda \int d \omega \; g(\omega) (e^{i \omega n_\lambda\ell} + (-1)^j )\ket{\lambda, \omega}$. The probability for the photon to be found in path $j$ can be computed as $p_j = \langle\phi_j|\phi_j\rangle$, resulting in
\begin{equation}
    \begin{split}
        p_j & = \frac{1}{4} \sum_\lambda |C_\lambda|^2\int d\omega |g(\omega)|^2 |e^{i \omega n_\lambda \ell} + (-1)^j|^2\\
        & = \frac{1}{4} \sum_\lambda |C_\lambda|^2 \bigg[2 + 2 \int d\omega |g(\omega)|^2 \cos{(\omega n_\lambda \ell)}\bigg].
    \end{split}
\end{equation}
By defining $\eta_\lambda = 2 \int d\omega |g(\omega)|^2 \cos{(\omega n_\lambda \ell)}$ we retrieve the probability as expressed in Eq.~(\ref{eq:path_prob}). The behavior of the probability for a gaussian frequency distribution is shown in Fig.~\ref{fig: prob}. The output state conditioned on the detection of a photon in a specific output branch $j$ can then be written as $\ket{\psi_{out, j}} = \frac{\ket{\phi_j}}{\sqrt{p_j}}$. Finally, the reduced polarization state is obtained by tracing out the frequency, $\rho_j = \text{Tr}_\omega (\ket{\psi_{out, j}}\bra{\psi_{out, j}})$, yielding
\begin{align}
    \rho_j = \frac{1}{4p_j}\begin{pmatrix}
(2 + (-1)^j \eta_H)|C_H|^2 & C_H C_V \kappa_j \\
C_H C_V \kappa_j^* & (2 + (-1)^j \eta_V)|C_V|^2\\
\end{pmatrix}. 
\end{align}

\begin{figure}
    \centering
    \includegraphics[width=0.99\linewidth]{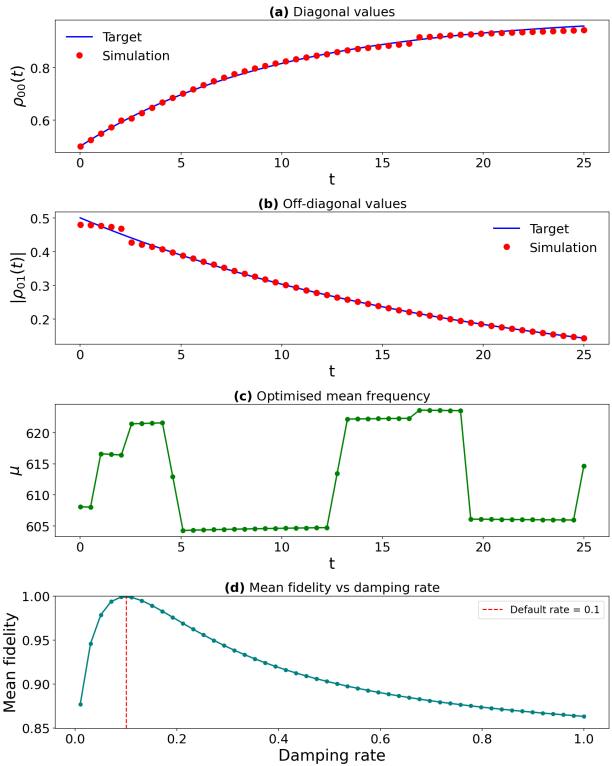}
    \caption{Results of the single qubit simulation for a amplitude damping dynamics, as described in Eq.~(\ref{eq: amp_damp dynamics}) \textbf{(a):} Comparison of the diagonal element $\rho_{00}$ \textbf{(b):} Comparison of the off-diagonal coherence \textbf{(c):} Values of the mean frequency $\mu$ in the unit of $THz$. Mean fidelity of the simulation shown as function of the depolarization rate $\gamma$. The interferometer parameters are separately optimized for each value of $\gamma$. For \textbf{(a)-(c)} the parameters are $\ell=0.467062 \cdot 10^{-12}~s$, $\sigma=0.538736 ~THz$, $n_H = 1.553$, $n_V = 1.544$, $|C_H| = 0.702459$, $\gamma_\downarrow = 0.1$. For \textbf{(d)} parameters $\ell$, $C_H$ and $\sigma$ are kept while a new set of $\mu$ values is separately optimized each $\gamma_\downarrow$.}
    \label{fig: amp_damp}
\end{figure}

\section{Non-Markovian evolution of a qubit in a lossy cavity}
\label{app: NME}
In Sec.~\ref{chap:3} of the main text, we have tested the single qubit simulation on an example of well-known Non-Markovian evolution. Here, we add more details on the target system. The Hamiltonian governing the system, under the R.W.A. is
\begin{equation}
    \begin{split}
            \hat{H} & = \frac{\omega}{2} \hat{\sigma}_z + \sum_k \omega_k \hat{a}^\dagger_k\hat{a}_k  +\sum_k (g_k\hat{\sigma}_+ \hat{a}_k + h.c.).
    \end{split}
\end{equation}
Here, $\hat{\sigma}_z = \{(1,0),(0,-1)\}$ represents a Pauli matrix while $\hat{\sigma}_+$ is the qubit ladder operator. $\hat{a}_k$ and $\hat{a}^\dagger_k$ are, respectively, the annihilation and creation operators for environment mode $k$. 
The decoherence function $G(t)$ can be obtained via the two point correlation function $f(t - s)$
\begin{equation}
    \frac{d}{dt}G(t) = - \int ds f(t - s) G(s).
\end{equation}
If the coupling parameters $g_k$ are distributed according to a Lorentzian distribution in the frequency space, $G(t)$ can be expressed as
\begin{equation}
    G(t) = e^{-\frac{\lambda}{2}} \left[ \cosh\left(\frac{dt}{2}\right) + \frac{\lambda}{d}\sinh\left(\frac{dt}{2}\right) \right],
\end{equation}
where $d  = \sqrt{\lambda^2 - 2 \lambda \gamma}$. From this, the time dependent Lamb shift $S(t)$ and the decay rate $\gamma(t)$ can be expressed as:
\begin{equation}
    \begin{split}
        & S(t) = -2 \mathcal{I}\{\dot{G}(t)/G(t)\}\\
        & \gamma(t) = -2 \mathcal{R}\{\dot{G}(t)/G(t)\}.
    \end{split}
\end{equation}
When $\lambda < 2\gamma$, $d$ has no real values and non-Markovianity can emerge in the form of coherence revivals \cite{breuer2016colloquium}.

\section{Extra example: Markovian amplitude damping}
\label{M-AD}
As an additional example to showcase the effectiveness of our protocol on a variety of commonly encountered noisy dynamics, we consider amplitude damping \cite{nielsen2010quantum}. This specific kind of noise is one of the fundamental single qubit decoherence models, describing energy relaxation. Physically, it models an irreversible loss of energy from the qubit to an environment and it characterizes processes like spontaneous emission from excited states.

For a continuous evolution in time, amplitude damping can be described as a Markovian Lindbladian dissipator
\begin{equation}
    \frac{d \rho}{dt} = \gamma_\downarrow \left(\hat{\sigma}_-\rho \hat{\sigma}_+ - \frac{1}{2} \{ \hat{\sigma}_+ \hat{\sigma}_-, \rho\}\right),
\end{equation}
where $\gamma_\downarrow$ is the inverse timescale for the noise process to take place. Specifically, this parameter represents how fast the excited state population decays into the lower energy state. Under this channel, any initial state decays into $\ket{0}$, which is the only stationary steady state for this evolution. Solving for a generic initial qubit density matrix yields the following time evolution.
\begin{equation} \label{eq: amp_damp dynamics}
    \rho (t) = \begin{pmatrix} 1 - \rho_{11}(t)e^{\gamma_{\downarrow}t}  & \rho_{01}(t)e^{\frac{\gamma_{\downarrow}}{2}t} \\ \rho_{10}(t)e^{\frac{\gamma_{\downarrow}}{2}t} & \rho_{11}(t)e^{\gamma_{\downarrow}t} \end{pmatrix}.
\end{equation}
In Fig.~\ref{fig: amp_damp}, we show results related to this case.

\end{document}